\documentclass[twocolumn,showpacs,superscriptaddress,amsmath,amssymb]{revtex4}
\usepackage{graphicx}
\usepackage{dcolumn}
\usepackage{bm}

\begin{document}

\title{Universal scaling in sports ranking}

\author{Weibing Deng} \affiliation{Complexity Science Center $\&$
Institute of Particle Physics, Hua-Zhong (Central China) Normal
University, Wuhan 430079, China} \affiliation{LUNAM Universit\'e,
ISMANS, Laboratoire de Physique Statistique et Syst\'emes
Complexes, 44, Ave. Bartholdi, 72000 Le Mans, France}
\affiliation{Universit\'e du Maine, LPEC, UMR CNRS 6087, 72085 Le
Mans, France}

\author{Wei Li}
\affiliation{Complexity Science Center $\&$ Institute of Particle
Physics, Hua-Zhong (Central China) Normal University, Wuhan
430079, China} \affiliation{Max-Planck Institute for Mathematics
in the Sciences, Inselstr. 22-26, D-04103 Leipzig, Germany}

\author{Xu Cai}
\affiliation{Complexity Science Center $\&$ Institute of Particle
Physics, Hua-Zhong (Central China) Normal University, Wuhan
430079, China}

\author{Alain Bulou}
\affiliation{Universit\'e du Maine, LPEC, UMR CNRS 6087, 72085 Le
Mans, France}

\author{Qiuping A. Wang}
\affiliation{LUNAM Universit\'e, ISMANS, Laboratoire de Physique
Statistique et Syst\'emes Complexes, 44, Ave. Bartholdi, 72000 Le
Mans, France} \affiliation{Universit\'e du Maine, LPEC, UMR CNRS
6087, 72085 Le Mans, France}

\begin{abstract}

Ranking is a ubiquitous phenomenon in the human society. By
clicking the web pages of Forbes, you may find all kinds of
rankings, such as world's most powerful people, world's richest
people, top-paid tennis stars, and so on and so forth. Herewith,
we study a specific kind, sports ranking systems in which players'
scores and prize money are calculated based on their performances
in attending various tournaments. A typical example is tennis. It
is found that the distributions of both scores and prize money
follow universal power laws, with exponents nearly identical for
most sports fields. In order to understand the origin of this
universal scaling we focus on the tennis ranking systems. By
checking the data we find that, for any pair of players, the
probability that the higher-ranked player will top the
lower-ranked opponent is proportional to the rank difference
between the pair. Such a dependence can be well fitted to a
sigmoidal function. By using this feature, we propose a simple toy
model which can simulate the competition of players in different
tournaments. The simulations yield results consistent with the
empirical findings. Extensive studies indicate the model is robust
with respect to the modifications of the minor parts.

\end{abstract}


\maketitle

\section{Introduction}

Most systems in nature have been perceived as a collection of a
huge amount of highly interacting units. Although equipped with
different components and interactions, different systems could
still possess some commonly shared characteristics. Through
analyzing the statistical distributions of varieties of empirical
quantities, two main patterns of distributions \cite{CSN} have
been found.

For the first one, there is a typical value, around which most
quantities distribute tightly clustered \cite{NewM}, or that is to
say, such pattern of distributions are peaked around this typical
value. Representative examples include the height of human beings,
the speed of cars on the motorway, the intelligence quotient (IQ)
test of people, etc. In the traditional IQ test for adults, most
people (about 69\% of the population) would score in the average
range (85-114), a small number (about 26\% of the population)
would score moderately below average (70-84) and moderately above
average (115-129), very high (130 or higher) and very low (below
70) are extremely rare (about 5\% of the population).

However, not all quantities could be well characterized by their
average values, some would change over an enormous dynamic range,
sometimes even many orders of magnitude. Such pattern has
absolutely long been familiar with, in the studies regarding the
distributions of people's annual incomes \cite{Pareto}, word
frequencies in text \cite{Zipf1}, and city sizes \cite{Zipf2}. For
the city sizes distribution, if cities are ranked by their
population from the largest (rank 1) to the smallest (rank N), it
is immediately discovered that, only a small number of cities
possess the large population, the majority of cities have the
small population. Relationship between rank and city sizes has
been found to follow $\ln rank = \alpha+\beta \ln size$, with the
slope of the curve $\beta$ being close to -1, which has been well
known as Zipf 's law.

A more general expression of such pattern is the power law
distribution, with $p(x)=Ax^{-\beta}$, $A=e^\alpha$, and $x$ is
the observation of the system. Power law distributions have been
considerably widely observed in nature, such as the net worth of
the richest individuals in the US \cite{NewM}, the frequencies of
occurrence of words in most human languages \cite{NewM,Zipf1,
Zipf2}, the frequencies of family names in most cultures
\cite{FamN}, the number of calls received by customs \cite{Abello,
Aiello}, the number of bytes of data received from computer users
\cite{Willinger}, the number of hits on the web sites
\cite{Adamic}, the number of links to web sites \cite{Broder}, the
number of citations received by papers \cite{Price}, the sizes of
computer files (such as the email address books) \cite{CME,NewM2},
the sizes of earthquakes \cite{Earthquake}, wars \cite{War},
craters on the moon \cite{Moon} and solar flares \cite{Solar}, the
severity of worldwide terrorist attacks \cite{Attack}, the number
of species per genus of mammals \cite{Species}, the number of
sightings of birds of different species \cite{Bird}, the sales of
books \cite{Book}, and music recordings \cite{Cox, Kohli}, etc.

As is well known, ranking is a very interesting and ubiquitous
phenomenon in the human society as every one tends to seek the
best. By clicking the web pages of Forbes you can find all kinds
of rankings, from world's most powerful people to world's richest
people, from top-paid models to American's top colleges, etc. Our
interest here is certainly not the gossip-like topic, but rather
whether there are some common patterns in the vastly different
ranking systems. Moreover, if yes, can we understand the formalism
of such patterns? To facilitate our study we choose a specific
kind of ranking systems, sports ranking, in which data are more
suitable for analysis. Here players' performance in attending
various competitions will be used as the basis of their respective
rankings, in terms of scores and/or prize money. Amazingly we find
that the distributions of scores and/or prize money follow
universal power-laws, with exponents being nearly identical for
different sports fields. The universal scalings can be reproduced
by our model in which the key mechanism is concerned with win-loss
probability distribution for any pair of players. This win-loss
probability distribution has been verified by the empirical data.
Our model is found to be robust with respect to the small
modifications of minor parts.

\section{Empirical results of sports ranking systems}

To understand how a certain sports ranking system works, let us
take tennis as an example. ATP (Association of Tennis
Professionals) and WTA (Women's Tennis Association) are world's
most successful tennis associations for male and female
professionals, respectively. To appear on the ranking systems of
ATP or WTA, the number of tournaments a player has to play should
reach a minimum, say 10. Tournaments have been divided into
several categories, such as grand slams, premier tournaments,
international tournaments and year-ending tour championships,
mainly based on the prize money. For the most important
tournaments such as grand slams, the main draw only consists of
128 players. The entry rule is that if you are top-ranked, then
you have more chances to attend the important tournaments. On the
other hand, players' good performance will improve their rankings
which will in turn entitle them more chances to play tournaments.
Since there are so many tournaments each year, for both ATP and
WTA, the ranking list of scores and of prize money vary from week
to week. Here we are not interested in which specific player is
world No.1 in certain sports, but instead the statistical
distribution of performance, measured by scores and prize money,
of all the member players. What is the form of such a
distribution? Is it stable over different time periods? Is it
universal?

Our data sets cover 12 different sports fields, such as tennis,
golf, snooker, and volleyball, etc. All the data are updated up to
February 2011. As the sample size of data is small, we adopt the
cumulative distribution to reduce the potential statistical
errors.

\subsection{Cumulative distribution of scores}

A player's score or prize money is a direct measure of his/her
performance in various competitions. The higher the score, the
better the performance. The statistical distribution of scores or
prize money reflects the profile of the performance of all the
members belonging to the same association. Every sports field has
its own scoring system, hence the orders of scores are not always
at the same level. In order to make the distributions of scores or
prize money comparable for different sports fields, we rescale the
quantities of interest. That is,
\begin{equation}
R_S=S/S_{max},
\end{equation}
where $S$ denotes the values of quantities considered, e.g.,
scores or prize money, and $S_{max}$ is the maximum value of $S$
in the sample, which pertains to the No. 1 player in the ranking
list by using $S$.

Cumulative distributions of players' scores or prize money have
been shown in Fig. 1 for 12 different sports ranking systems.
Amazingly all the distributions share very similar trend, which
can be well fitted to the power-law with an exponential cutoff as
below,
\begin{equation}
P_>(S)\propto S^{-\tau}exp(-S/S_c),
\end{equation}
where $\tau$ and $S_c$ are critical exponent and size cutoff,
respectively. It should also be noticed that for the same field,
all the curves collapse with each other. Values of $\tau$ and
$S_c$ for different sports fields are given in Table 1, where
values of $\tau$ range from 0.01 to 0.39, and the counterparts of
$S_c$, from 0.12 to 0.28.

\begin{table}
\centering \caption{System sizes of 40 samples in the 12 different
sports ranking systems, values of the critical exponent $\tau$ and
size cutoff $S_c$ in the power law with exponential cutoff, {\it
p-values} for the statistical significance test, and the ratio of
the Pareto principle test.}
\begin{tabular*}{\hsize}{@{\extracolsep{\fill}}lcrrrr}
Sports ranking systems& Sizes &$\tau$& $S_c$ & $p$\tablenote{{\it
P-value} of Kolmogorov-Smirnov (KS) test for the cumulative scores
or prize money distributions, with hypothesized distribution being
the power law with exponential cutoff.}& ratio\tablenote{Values of
the ratio for the test of Pareto principle.}\cr \hline ATP Single&
1763&0.31&0.12& 0.65& 0.79\cr ATP Double& 1516&0.32&0.18&
0.52&0.78\cr ATP Prize Money&1636&0.33&0.13& 0.56&0.79\cr WTA
Single&1523&0.39&0.15& 0.62&0.78\cr WTA Double&1028&0.38&0.19&
0.75&0.80\cr WTA Prize Money&1388&0.39&0.12& 0.81&0.81\cr PGA
Score  &1323&0.16&0.18& 0.85&0.82\cr LPGA Score&734&0.18&0.19&
0.82&0.78\cr PGA Average Score&1323&0.16&0.19& 0.76&0.79\cr LPGA
Average Score &734&0.17&0.20& 0.82&0.82\cr ITTF Prize Money
Men&1717&0.32&0.17& 0.85& 0.83\cr ITTF Prize Money
Women&1288&0.32&0.18& 0.73&0.82\cr FIVA Junior Men&105&0.16&0.21&
0.86&0.76\cr FIVA Junior Women&95&0.14&0.20& 0.68&0.79\cr FIVA
Senior Men&138 &0.13&0.16& 0.69&0.78\cr FIVA Senior
Women&127&0.11&0.18& 0.92&0.82\cr FIFA Men&209&0.01&0.19&
0.59&0.77\cr WPBSA Total Score&97&0.11&0.27& 0.69&0.83\cr WPBSA
Average Score&97&0.13&0.25& 0.58&0.78\cr BWF Women
Single&548&0.12&0.16& 0.68&0.80\cr BWF Women Double&295&0.13&0.18&
0.53&0.78\cr BWF Men Single&833&0.06&0.17& 0.62&0.82\cr BWF Men
Double&429&0.08&0.13& 0.75&0.81\cr BWF Mixed Double&407&0.07&0.14&
0.63&0.79\cr FIBA Men&79&0.19&0.20& 0.86&0.81\cr FIBA Women&72
&0.18&0.21& 0.98&0.83\cr FIBA Boys&77&0.18&0.23& 0.62&0.82\cr FIBA
Girls& 72&0.26&0.22& 0.85&0.76\cr FIBA
Combined&115&0.23&0.20&0.52&0.81\cr IBAF Men& 78&
0.20&0.28&0.96&0.79\cr FIH Men& 73 &0.23&0.26&0.86&0.78\cr FIH
Women & 68&0.21&0.27&0.83&0.81\cr IHF Men &52& 0.16 &
0.25&0.68&0.79 \cr IHF Women & 46 & 0.15& 0.27 & 0.69 &0.76 \cr
FIE Sabre Senior Women& 371&0.34 & 0.25&0.56 &0.81 \cr FIE Foil
Senior Women&260 &0.32 &0.23 &0.65&0.78 \cr FIE Epee Senior Women
&293 &0.36 & 0.24& 0.53&0.83\cr FIE Sabre Senior Men & 319 &0.32 &
0.23&0.67&0.78 \cr FIE Foil Senior Men & 337&0.30 &0.21 &0.56
&0.82\cr FIE Epee Senior Men & 442&0.28 &0.25 &0.72 & 0.81
\cr\hline
\end{tabular*}
\end{table}

We employ the Kolmogorov-Smirnov (KS) test to quantify how closely
the power laws with exponential cutoffs resemble the actual
distributions of the observed sets of samples. Based on the
observed goodness of fit, the {\it p-value}, which is defined to
be the probability that the real data are drawn from the
hypothesized distribution, is calculated for each set of sample.
The {\it p-values} given in Table 1 for the statistical
significance test are all much larger than 0.1, thereby we could
conclude the power laws with exponential cutoffs are reliable fits
to the samples of different sports ranking systems.

The evidence of the power-laws in the sports ranking indicates
that there is still significant probability to have superman such
as Roger Federer in tennis or Tiger Woods in golf. But the
prevalent probability is still the players who do not play in the
top form. Unlike the human height system, it seems there is no
typical player who plays with average level. The power-laws found
here are also different from Zipf's law in which the critical
exponent is -1, much larger than ours (in absolute value).

\begin{figure*}[ht]
\begin{center}
\centerline{\includegraphics[width=0.9\textwidth]{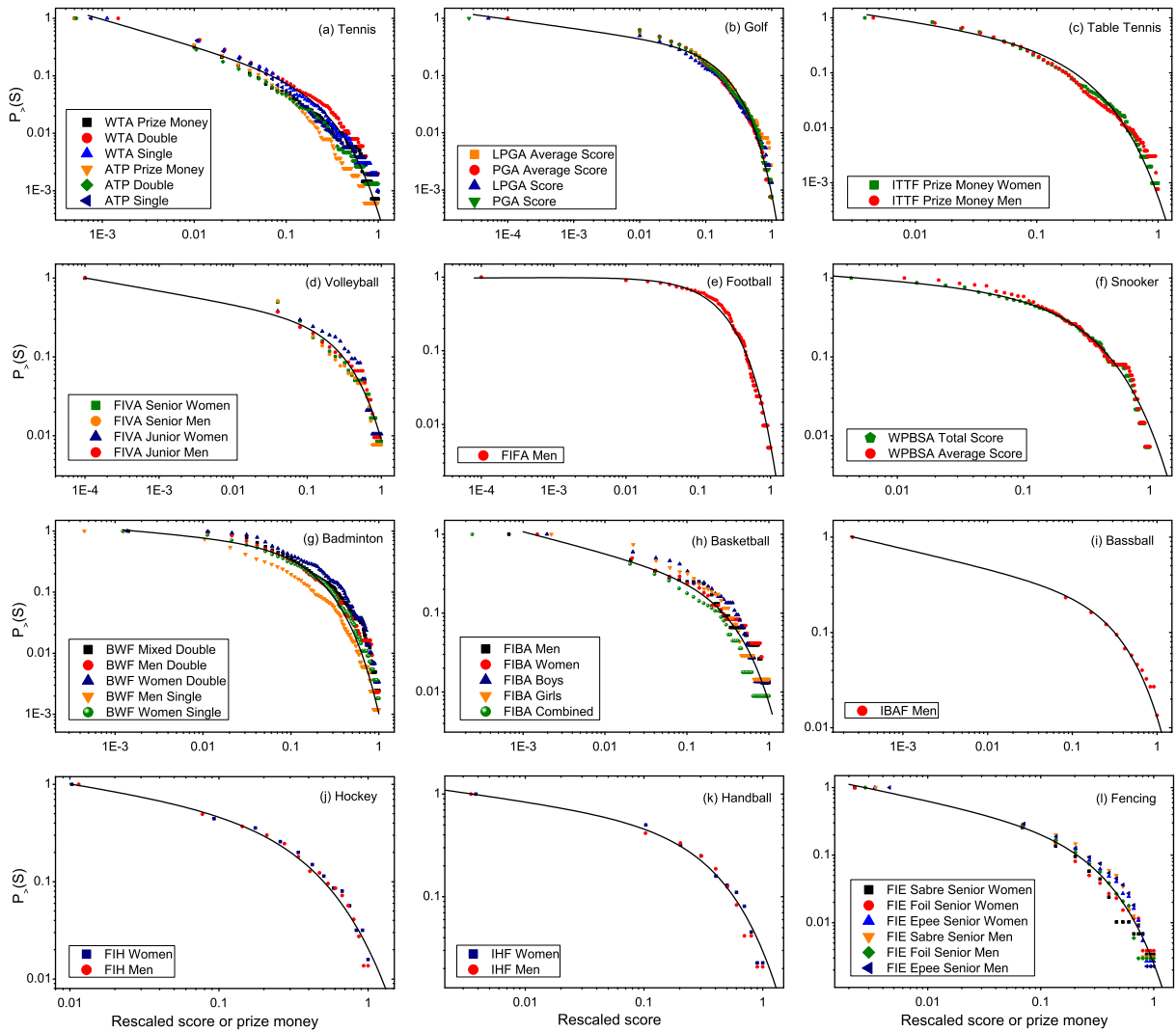}}
\caption{Cumulative distributions of scores and prize money for 12
different sports fields. (a) Tennis: Association of Tennis
Professionals (ATP) and Women's Tennis Association (WTA). (b)
Golf: Professional Golfers' Association (PGA) and Ladies
Professional Golf Association (LPGA). (c) Table tennis:
International Table Tennis Federation (ITTF). (d) Volleyball:
International Federation of Volleyball (FIVB). (e) Football:
International Federation of Football Association, commonly known
as FIFA. (f) Snooker: World Professional Billiards and Snooker
Association (WPBSA). (g) Badminton: Badminton World Federation
(BWF). (h) Basketball: International Basketball Federation, more
commonly known as FIBA. (i) Baseball: International Baseball
Federation (IBAF). (j) Hockey: International Field Hockey
Federation (FIH). (k) Handball: International Handball Federation
(IHF). (l) Fencing: International Fencing Federation (FIE). All
the curves can be fitted to power laws with exponential cutoff,
$P_>(S)\propto S^{-\tau}exp(-S/S_c)$, where $\tau$ is the critical
exponent and $S_c$ is the size cutoff (the turning point). The
values of $\tau$ and $S_c$ for different sports fields are
provided in Table 1.}\label{afoto2}
\end{center}
\end{figure*}

\subsection{Pareto principle}

The Pareto principle \cite{Pareto2}, also well known as the 80-20
rule, states that, for many events, roughly 80\% of the effects
comes from 20\% of the causes. Pareto noticed that, 80\% of
Italy's land was owned by 20\% of the population. He carried out
such surveys on a variety of other countries further, and to his
surprise, the rule was also fulfilled.

The 80-20 rule has also been used to attribute the widening
economic inequality, which showed that, the distribution of global
income to be very uneven, with the richest 20\% of the world's
population controlling 82.7\% of the world's income. The 80-20
rule could be applied to many systems, from the science of
management to the physical world.

We also check this rule in the sports ranking systems, it is
interesting to find that, 20\% players indeed possess
approximately 80\% scores or prize money of the whole system, the
ratios we got in different sports ranking systems are shown in
Table. 1, values of the ratios are all very close to 0.8.

\subsection{Dependence of win probability on $\Delta$ rank}

Here we employ the concept of "win probability" to describe the
chances that a player or a team will win when encountering an
opponent. For instance, what is the odds that a No.1 player will
top a No.100 player? What is again her chance against No.2?
Theoretically, the chance is much higher in the former case than
in the latter one. But the result of a competition is not unknown
until it is over, which mainly depends on how the player performs
at that specific match. However, the win probability could be
solely based on the previous performance of a player against a
certain opponent, which then can used to predict her future
performance against the same opponent. This might have some
applications in betting the result of a match. To simplify the
case without loss of generality, we relate the win probability
solely to the rank difference of a pair of players. Suppose we now
have two players A and B, with A having a higher rank. We will
then need to know how likely A can beat B when they meet? This
quantity is related to but different from the win percentage we
usually refer to. The win percentage depicts the percentage of win
of a player over all previous encounters. We assume that the win
probability only depends on the rank difference between two
players. This means, the probability that No.1 beats No.100 is the
same as the one that No.100 beats No.200. Hence, we have the
following definition,
\begin{equation}
P_{win}(\Delta r)=\frac{N_{win}(\Delta r)}{N_{total}(\Delta r)},
\end{equation}
where $\Delta r$ denotes the rank difference (integer), $N_{win}
(\Delta r)$ is the total number of win for the higher-ranked
players when the rank difference is $\Delta r$, and
$N_{total}(\Delta r)$ is the total number of matches in which the
rank difference between the pair is $\Delta r$. We here emphasize
again that the win probability is the probability that the
higher-ranked player will win when two players meet. When $\Delta
r$ is small, say 1, it is difficult to judge which player will
win, and in this case $P_{win}$ might approximately equal 0.5.
When $\Delta r$ is large, for instance 100, $P_{win}$ might
approach 1, which means the higher-ranked player is very likely to
win.

By using the {\it Head to Head} records of ATP and WTA, we find
that the dependence of $P_{win}$ on $\Delta r$ can be well fitted
to the sigmoidal function as follows,
\begin{equation}
P_{win}=\frac{1}{1+exp{(-a*\Delta r)}},
\end{equation}
where $a$ is a parameter dependent on the specific systems. For
ATP and WTA, $a$ is 0.021 and $0.032$, respectively. The existence
of fluctuations is quite natural since even Roger Federer will not
win all the matches. The value of $a$ can still tell us some
information about how competitive that certain sports is. The
smaller $a$ is, the more competitive the sports will be. Let us
take WTA and ATP as two examples. When $\Delta r$ is 30, the win
probability for WTA is nearly 0.7, while the counterpart for ATP
is 0.65. This means the game is more unpredictable in ATP than in
WTA. It is not strange since men's game is more competitive than
women's. We of course wish to test the empirical finding by
checking data from other sports fields, but not so many data are
available as far as we know. Despite this, this finding will play
a key role in our model to follow.

\begin{figure}[ht]
\begin{center}
\centerline{\includegraphics[width=0.35\textwidth]{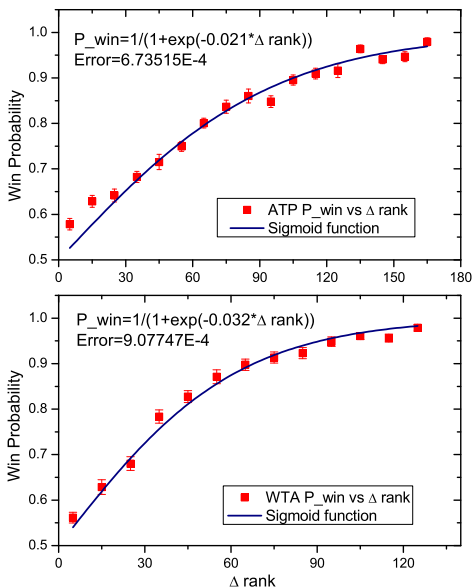}}
\caption{Dependence of win probability on $\Delta r$ of the
players for ATP and WTA, which can be well fitted to the sigmoidal
function $P_{win}=1/(1+exp{(-a*\Delta r)})$.}\label{afoto2}
\end{center}
\end{figure}

\section{A simple toy model of sports ranking systems}

What is the origin of the universal scaling in different sports
systems? Of course, there have been so many approaches which can
explain the origin of power-laws. Some mechanisms or theories are
elegant, e.g., random walks \cite{NewM}, self-organized
criticality (SOC) \cite{Bak}, etc. It is, however, difficult to
try to apply these frameworks to sports ranking systems. We
propose a simple toy model, inspired by tennis. Of course, the
model may not suit any sports field but does have some general
implications. Most importantly, our model can reproduce robust
power-laws without having to introduce additional parameters.

\begin{figure}[ht]
\begin{center}
\centerline{\includegraphics[width=0.35\textwidth]{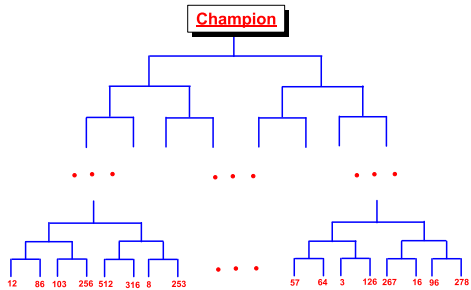}}
\caption{A cartoon of a draw sample. After each round, half
players will be eliminated, the numbers "12, 86 ..." denote the
ranks of the players.}\label{afoto2}
\end{center}
\end{figure}

The rules of the model are defined in the following way,

(1) $2^N$ players are ranked from 1 to $2^N$, being assigned
random scores drawn from a Gaussian distribution.

(2) For each tournament, all the players have entry permission.
Therefore the draw will include $2^N$ players and in total $N$
rounds. At each round, half of the players will be eliminated when
they lose. The rest will enter the next round. The losers at round
$n$ will gain score $2^{(n-1)}$. The final champion wins score
$2^N$.

(3) The key mechanism is to decide which one will lose for a given
pair of players. Here our empirical finding will be employed.
Namely, when two players meet, the probability that the
higher-ranked player will top the lower-ranked opponent is given
by $1/(1+exp(-a*\Delta r))$, where $\Delta r$ is their rank
difference, as before.

(4) A new tournament opens up and a new draw is made.

In principle, there is only one parameter in our model, that is
$a$. We can simply call it competition strength. Of course, there
are some shortcomings in the model. First, in the actual
tournaments not all the players will be accepted. In grand slams
there are only 128 players. Second, tournaments can be divided
into many categories and may consist of different players. Third,
the scoring systems for different tournaments are a little
different. For grand slams the scores and prize money are much
higher than other tournaments, if the players are eliminated at
the same round. We certainly can add these issues into our model
in order to test the resilience of the model. At the moment we do
not wish to complicate the model by introducing additional
parameters. What we need here is a skeleton which may allow us to
understand some key features of the specific systems. Namely, if
the power-laws with exponential cutoffs can be reproduced through
our model, then it is a feasible model. We need not to care about
other minor issues.

\begin{figure}[ht]
\begin{center}
\centerline{\includegraphics[width=0.35\textwidth]{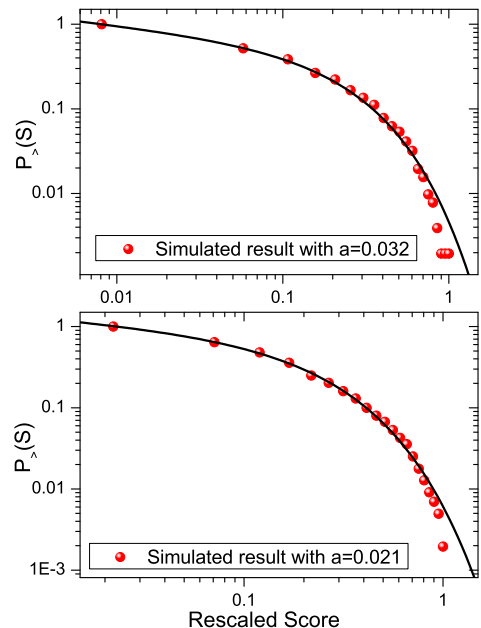}}
\caption{Cumulative distribution of scores from the simulation.
For these two samples, number of players $N_p=2048$, and number of
total tournaments $N_t=128$, with $P_{win}=1/(1+exp(-a*\Delta
r))$, $a=0.032$ and $0.021$, respectively.}\label{afoto2}
\end{center}
\end{figure}

\section{Simulation results and discussions}

The most important parameter in our model is $a$, the so-called
competition strength. The number of players $N_p$ and the number
of tournaments $N_t$ only have finite-size effects. It is natural
to check the dependence of the simulation results on these
parameters, which can reflect the resilience of our model.

First of all, we need to test whether the model can reproduce the
power-laws of the cumulative distribution of scores. In Fig. 4,
$N_p$ equals 2048, and $N_t$ is 128, while win probability,
$P_{win}=1/(1+exp(-a*\Delta r))$, with $a=0.021$ and $0.032$, as
given by the empirical data of ATP and WTA, respectively. We find
that, the cumulative distributions of scores given by the
simulations indeed follow the power-law distributions with
exponential cut-off, $P(S)\propto S^{-\tau}exp(-S/S_{c})$, with
$\tau=0.2, 0.22$, $S_c=0.23, 0.19$, respectively for these two
samples. Here we notice that the values of the parameters are very
close to what are obtained from the experimental data.

\begin{figure}[ht]
\begin{center}
\centerline{\includegraphics[width=0.35\textwidth]{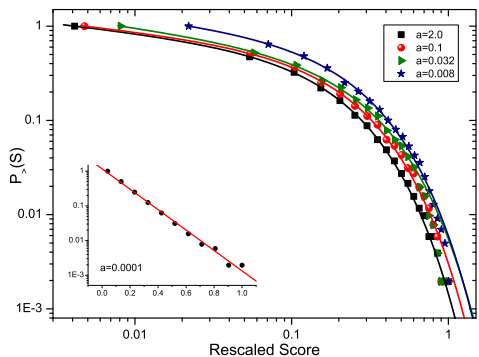}}
\caption{Influence of the critical parameter $a$ on the final
cumulative scores distributions, values of $a$ ranging from 0.0001
to 2.0.}\label{afoto2}
\end{center}
\end{figure}

In the formula of win probability, smaller values of $a$
correspond to more intensive competition. For instance, when
$a=0.0001$, $P_{win} \leqslant 0.525$ for $\Delta r \leqslant
1000$, which means higher-ranked player only has slightly more
chances than the lower-ranked player to win the match between
them. While larger values of $a$ suggest that the higher ranked
players would win the match with a much larger probability. For
example, when $a=2.0$, $P_{win} \geqslant 0.88$ for $ \Delta rank
\geqslant 1$.

\begin{figure}[ht]
\begin{center}
\centerline{\includegraphics[width=0.35\textwidth]{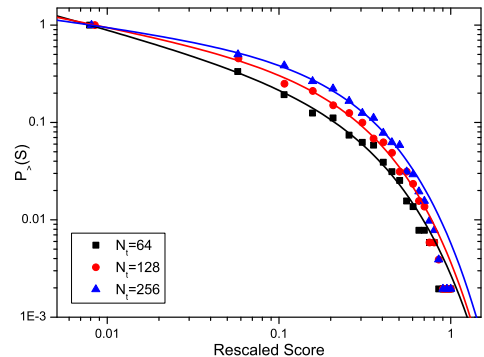}}
\caption{Simulation result of the cumulative scores distributions
for different number of tournaments, with $N_t$=64, 128, and
256.}\label{afoto2}
\end{center}
\end{figure}

Thus here, to analyze the influence of win probability, we
simulated our models with different values of $a$,
$0.0001\leqslant a \leqslant 2.0$. From Fig. 5, we can find that
the cumulative scores distributions change from the power laws
with exponential cutoff to exponential. Since when $a$ is very
small, such as $a=0.0001$, all players nearly win the match
randomly, thus the cumulative probabilities of the scores
approximately like $1$, $1/2$, $(1/2)^2$, ..., which results the
exponential format.

For different number of tournaments, $N_t= 64, 128$ and $256$, the
cumulative distributions of scores are shown in Fig. 6, as seen,
one could discover, all the cumulative distributions of scores are
power- laws with exponential cutoff, values of the critical
exponents $\tau$ and size cutoff $S_c$ are also very close to
those of the empirical results.

In statistical physics, in order to determine the validity of the
statistical approach, we often take the thermodynamic limit, in
which the number of components $N$ tends to infinity
\cite{Finite1}. However, in real world networks, the number of
vertices or agents can never be that large, this makes the factor
of finite size of paramount importance. For example, even the
largest artificial net, the World Wide Web, whose size will soon
approach $10^{11}$ Web pages, also shows qualitatively strong
finite-size effects \cite{Finite2}.

\begin{figure}[ht]
\begin{center}
\centerline{\includegraphics[width=0.35\textwidth]{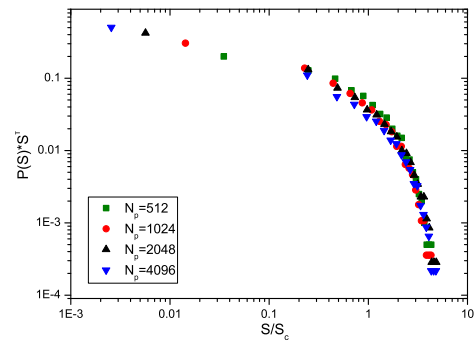}}
\caption{Finite size effects analysis of the simulation results,
$N_p$=512, 1024, 2048 and 4096. }\label{afoto2}
\end{center}
\end{figure}

Therefore, here, in order to test the influence of the finite size
effect on the final cumulative scores distribution, we considered
the transformed score distribution $P(S)*S^\tau$ versus $S/S_c$,
where $S_c$ is the characteristic size cutoff. For four different
system sizes, such relationships were shown in Fig. 7, which
suggest that, the tails of the four curves almost collapsed with
each other, thereby, we can conclude the finite size effect is
almost negligible.

\section{Conclusion}

In summary, to characterize the intrinsic common features and
underline dynamics of ranking systems, we carry out the
investigations in an applicable and specific kind of ranking
systems, the sports ranking systems, the main results are: (i) The
universal scaling law is extensively found in the distributions of
scores and/or prize money, in addition, values of the critical
exponents are close to each other for 40 samples of 12 sports
ranking systems. (ii) Players' scores are discovered to obey the
Pareto principle, which means, 20\% of players approximately
possess 80\% of total scores of the whole system. (iii) Win
probability is introduced to describe the chance that a player or
a team will win when meeting an opponent, we simply relate the win
probability solely to the rank difference $\Delta r$, for tennis
sport, the win probability has been empirically verified to follow
the sigmoid function, $P_{win}=1/(1+exp{(-a*\Delta r)})$. (iv) By
employing the empirical features of win probability, we proposed a
simple toy model to simulate the real process of the sports
systems, the universal scaling could be well reproduced by our
model, moreover, this result is robust when we change the values
of parameters in the model.

We are expecting to find such similar scaling laws in other
ranking systems, and we hope all these results and methods could
be well applied to analyzing any type of paired competitions, or
solving some practical problems in the ranking systems.

\section*{Acknowledgement}

W.B. Deng would like to show gratitude to Cyril Pujos, Laurent
Nivanen for fruitful discussions. This work was supported by
National Natural Science Foundation of China (Grant Nos.10647125,
10635020, 10975057 and 10975062), the Programme of Introducing
Talents of Discipline to Universities under Grant No. B08033, and
the PHC CAI YUAN PEI Programme (LIU JIN OU [2010] No. 6050) under
Grant No. 2010008104.


\begin{thebibliography}{10}

\bibitem{CSN}
Clauset A, Shalizi CR, and Newman MEJ (2009) Power-law
distributions in empirical data. {\it SIAM Review} 51(4): 661-703.

\bibitem{NewM}

Newman MEJ (2005) Power laws, Pareto distributions and Zipf's law.
{\it Contemporary Physics} 46(5): 323-351.

\bibitem{Pareto}

Pareto V (1896)  Cours d'Economie Politique. {\it Gen\`eve: Droz.}

\bibitem{Zipf1}

Zipf GK (1932) Selected Studies of the Principle of Relative
Frequency in Language. {\it Cambridge, MA.: Harvard University
Press.}

\bibitem{Zipf2}

Zipf GK (1949) Human Behavior and the Principle of Least Effort.
{\it Cambridge, MA: Addison-Wesley.}

\bibitem{FamN}

Zanette DH and Manrubia SC (2001) Vertical transmission of culture
and the distribution of family names. {\it Physica A} 295: 1-8.

\bibitem{Abello}

Abello J, Buchsbaum A and Westbrook J (1998) A functional approach
to external memory graph algorithms. {\it Proceedings of the 6th
European Symposium on Algorithms} 1461: 332-343.


\bibitem{Aiello}

Aiello W, Chung F. and Lu L (2000) A random graph model for
massive graphs. {\it Proceeding of the 32nd Annual ACM Symposium
on Theory of Computing} 171-180.

\bibitem{Willinger}

Willinger W and Paxson V (1998) Where Mathematics meets the
Internet. {\it Notices of the American Mathematical Society} 45:
961.

\bibitem{Adamic}

Adamic LA and Huberman BA (2000) The nature of markets in the
World Wide Web. {\it Quarterly Journal of Electronic Commerce} 1:
512.

\bibitem{Broder}

Broder A, Kumar R, Maghoul F, Raghavan P, Rajagopalan S, Stata R,
Tomkins A and Wiener J (2000) Graph structure in the Web. {\it
Computer Networks} 33: 309.


\bibitem{Price}

Price DJdeS (1965) Networks of scientific papers. {\it Science}
149: 510-515.

\bibitem{CME}

Crovella ME and Bestavros A (1996) Self-similarity in World Wide
Web traffic: Evidence and possible causes. {\it Proceedings of the
1996 ACM SIGMETRICS Conference on Measurement and Modeling of
Computer Systems} 148-159.

\bibitem{NewM2}

Newman MEJ, Forrest S and Balthrop (2002) Email networks and the
spread of computer viruses. {\it Phys. Rev. E} 66: 035101


\bibitem{Earthquake}

Gutenberg B and Richter RF (1944) Frequency of earthquakes in
california. {\it Bulletin of the Seismological Society of America}
34: 185-188.

\bibitem{War}

Roberts DC and Turcotte DL (1998) Fractality and selforganized
criticality of wars. {\it Fractals} 6: 351-357.

\bibitem{Moon}

Neukum G and Ivanov BA (1994) Crater size distributions and impact
probabilities on Earth from lunar, terrestial-planet, and asteroid
cratering data. {\it Hazards Due to Comets and Asteroids,
University of Arizona Press, Tucson, AZ} 359-416

\bibitem{Solar}

Lu ET and Hamilton RJ (1991) Avalanches of the distribution of
solar flares. {\it Astrophysical Journal} 380: 89-92.

\bibitem{Attack}
Clauset A, Young M and Gleditsch KS (2007) On the Frequency of
Severe Terrorist Events. {\it Journal of Conflict Resolution} 51:
58.


\bibitem{Species}

Willis JC and Yule GU (1922) Some statistics of evolution and
geographical distribution in plants and animals, and their
significance. {\it Nature} 109: 177-179.

\bibitem{Bird}

{\it North American Breeding Bird Survey, Results and Analysis},
1966-2003.

\bibitem{Book}

Hackett AP (1967) 70 Years of Best Sellers, 1895-1965. R. R.
Bowker Company, New York.


\bibitem{Cox}

Cox RAK, Felton JM, and Chung KC (1995) The concentration of
commercial success in popular music: an analysis of the
distribution of gold records. {\it Journal of Cultural Economics}
19: 333-340.

\bibitem{Kohli}

Kohli R and Sah R (2003) Market shares: Some power law results and
observations. {\it Working paper 04.01, Harris School of Public
Policy, University of Chicago.}

\bibitem{Pareto2}
Pareto, Vilfredo; Page, Alfred N. (1971), Translation of Manuale
di economia politica ("Manual of political economy"), A.M. Kelley,
ISBN 9780678008812

\bibitem{Bak} Bak P, Tang C, and Wiesenfeld K (1987), Self-organized
criticality: An explanation of the 1/f noise. {\it Physical Review
Letters} 59: 381-384.

\bibitem{Finite1}

Toral R and Tessone CJ (2007) Finite size effects in the dynamics
of opinnion formation. {\it Commun. Comput. Phys.} 2: 177-195.

\bibitem{Finite2}

Dorogovtsev SN, Goltsev AV and Mendes JFF (2008) Critical
phenomena in complex networks. {\it Rev. Mod. Phys.} 80:
1275-1335.


\end{thebibliography}
\end{document}